\documentclass{llncs}

\usepackage{lineno,hyperref}

\usepackage{amsmath}
\usepackage{amssymb}
\usepackage{xspace}
\usepackage{proof-dashed}
\usepackage{tikz}
\usepackage[arrow,matrix,curve]{xy}
\usepackage{float}
\usepackage[caption = false]{subfig}

\usepackage{graphicx}
\usepackage{grffile}

\usepackage{booktabs} 
\usepackage{float}
\usepackage{listings}
\usepackage{todonotes}
\usepackage{mathtools}
\DeclarePairedDelimiter{\ceil}{\lceil}{\rceil}
\newcommand{\vcpu}{\textsf{vCPU}}

\title{Microservice Dynamic Architecture-Level Deployment Orchestration (Extended Version)}
\author{Lorenzo Bacchiani\inst{1} \ \ \ Mario Bravetti\inst{1,2} \ \ \ Saverio Giallorenzo\inst{1,2} \\ \ \ \ Jacopo Mauro\inst{3} \ \ \ Iacopo Talevi\inst{1} \ \ \ Gianluigi Zavattaro\inst{1,2}}
\institute{Universit\`a di Bologna, Italy \and
Focus Team, INRIA, France \and 
University of Southern Denmark, Denmark
}

\date{}

\begin{document}

\maketitle

\begin{abstract}
In the context of the BI-REX (Big Data Innovation and Research Excellence) competence center SEAWALL (SEAmless loW lAtency cLoud pLatforms) project (scientific coordinator Prof. Maurizio Gabbrielli) we develop a novel approach for run-time global adaptation of microservice applications, based on synthesis of architecture-level reconfiguration orchestrations.  More precisely, we devise an algorithm for automatic reconfiguration that reaches a target system Maximum Computational Load by performing optimal deployment orchestrations.  To conceive and simulate our approach, we introduce a novel integrated timed architectural modeling/execution language based on an extension of the actor-based object-oriented Abstract Behavioral Specification (ABS) language.  
In particular, we realize a timed extension of SmartDeployer, whose ABS code annotations 
make it possible to express architectural properties. 
Our Timed SmartDeployer tool fully integrates time features of ABS and architectural annotations by generating timed deployment orchestrations.
We evaluate the applicability of our approach on a realistic microservice application taken from the literature: an Email Pipeline Processing System. We prove its effectiveness by simulating such an application and by comparing architecture-level reconfiguration with traditional local scaling techniques (which detect scaling needs and enact replications at the level of single microservices). Our comparison results show that our approach avoids cascading slowdowns and consequent increased message loss and latency, which affect traditional local scaling.
\end{abstract}

\section{Introduction}

Inspired by service-oriented computing, microservices structure software
applications as highly modular and scalable compositions of fine-grained and
loosely-coupled services~\cite{DGLMMMS17,BravettiZ09}. These features support modern
software engineering practices, like continuous
delivery/deployment~\cite{continuous_delivery} and application
autoscaling~\cite{autoscaling}. A significant problem in these practices consists
of the automated deployment of the microservice application: optimal
distribution of the fine-grained components over the available Virtual Machines (VMs), and dynamic reconfiguration to cope, e.g., with positive or negative
peaks of user requests.

Although these practices are already beneficial, they can be further improved 
by exploiting the interdependencies within an architecture (interface functional 
dependences), instead of focusing on the single microservice. 
Indeed, w.r.t.\ traditional local scaling techniques,
architecture-level dynamic deployment orchestration can:
\begin{itemize}
\item Avoid “domino” effects of unstructured scaling, i.e.\ single services scaling 
one after the other (cascading slowdowns) due to local workload monitoring.  
\item Quickly restore an acceptable performance in terms of message loss and latency.
\end{itemize}


In this paper, we first introduce a novel {\it integrated timed architectural modeling/execution language} based on an extension of the actor-based object-oriented Abstract Behavioral Specification (ABS) language \cite{abs_docs}. The extension that we devise crucially exploits the double nature of ABS, which is both a process algebra (it has a probabilistic/timed formal semantics) and a programming language (it is compiled and executed, e.g.\ with the Erlang backend). 
In particular, we realize a {\it timed} extension of SmartDeployer \cite{fase_paper,chapter}, whose ABS code {\it annotations} 
make it possible to express: {\it architectural properties} of the modeled distributed system (global architectural invariants and allowed reconfigurations), of its VMs (their characteristics and the resource they provide) and of its software components/services (their resource/functional requirements). Such annotations are read by SmartDeployer that, at compile-time, checks them for satisfiability (accounting for requirements and architectural invariants) and synthesizes deployment orchestrations that build the system architecture and each of its specified reconfigurations. SmartDeployer generates optimal deployment and undeployment code by using ABS itself as an orchestration language and by making it available via methods with conventional names. Such methods 
can be invoked by the ABS code of services, thus realizing run-time adaptation. Here we introduce the {\it Timed SmartDeployer tool} that fully realizes the integration between timed ABS execution language and architectural annotations by  generating {\it timed deployment orchestrations}. Such orchestrations also manage time aspects, dynamically setting 
VM speeds (based on virtual cpu cores that are actually being used) and overall startup time for the deployed architectural reconfiguration.

One of our main motivations in having a model encompassing architectural invariants/reconfigurations is to anticipate at the modeling level deployment orchestration related issues. This indeed fosters an approach where analysis of the consequences of deployment decisions are available early on: Timed SmartDeployer checks (at compile-time) the synthesizability of deployment orchestrations that, at run-time, will ensure the system to be always capable of adapting in case of positive/negative peaks of user requests.
On the contrary run-time deployment decisions, if left to loosely-coupled reactive scaling policies, could lead to a chaotic behavior in the system.

Moreover, in this paper we contribute an algorithm for architecture-level run-time adaptation that overcomes the 
shortcomings 
of the traditional local scaling approach.
We could conceive and simulate it thanks to the above architectural modeling/execution language. Such an algorithm finds application in the context of cloud-computing platforms endowed with orchestration engines.
The algorithm reaches, by
performing global reconfigurations, a target system Maximum Computational Load (MCL), i.e.\ the maximum supported frequency for inbound 
requests. 
The idea is that, by monitoring at run-time the inbound workload, 
our algorithm causes the system to be always in the reachable configuration that better fits such a workload (and that has the minimum number of deployed microservice instances). 
%
%
In particular, global reconfigurations 
are targeted at
guaranteeing a given increment (or decrement) of the system MCL. 
Moreover, we show how such an overall system MCL can be computed by the 
MCL of single service instances. In turn, they are
mathematically calculated based on: the microservice data rate (we use, e.g., real data in \cite{nginx} for Nginx servers) and the role it plays in the application architecture (which determines the mean number and size of its requests for each incoming message). 
As we will see, the timed features of deployment orchestrations synthesized by our Timed SmartDeployer tool are essential to model, in an MCL consistent way, adaptation actions enacted by our algorithm 
(dynamic speed of VMs and their overall startup time).

Finally, we evaluate the applicability of our approach on a realistic microservice application: an Email Pipeline Processing System taken from Iron.io~\cite{ironIO}.
Its model is built by considering: static aspects of the architecture (annotations) and ABS code modeling the behavior of services. 
We simulate system execution using inbound traffic inspired to two different real datasets in \cite{realWorkload} and \cite{KaramollahiW19}, representing the frequency of emails entering the system.
In order to show the effectiveness of our architecture-level adaptation algorithm, we compare it with traditional local scaling techniques. In particular, we produce two ABS programs: one implementing our algorithm (using $4$ Timed SmartDeployer synthesized orchestrations) and one just dealing with scaling needs at the level of single microservices.  Our comparison results show that our algorithm actually avoids cascading slowdowns and consequent increased message loss and latency that affect traditional local scaling. The obtained code fully exploits the expressive power of ABS, e.g.\ using both its timed and probabilistic features.\footnote{Complexity of our ABS process algebraic models is also witnessed by the fact that they led us to discover an error in the Erlang backend: it caused interferences in time evolution between unrelated VMs (it was solved thanks to our code).}

Wrapping up the novel contributions of this paper (e.g.\ compared to our previous work in \cite{fase_paper,chapter}) are: $(i)$ a novel integrated timed architectural modeling/execution language based on a timed extension of SmartDeployer that, differently from the previous version, exploits timed instructions of ABS to automatically generate timed deployment orchestrations, $(ii)$ an architecture-level run-time adaptation algorithm that reaches any target system MCL, $(iii)$ mathematical calculation of service MCL and MCL-based scaling configurations and $(iv)$ ABS code implementing system service execution/scaling mechanism for the Email Pipeline Processing System~\cite{ironIO}.

The paper is structured as follows. In section $2$ we briefly recall the microservice model, the ABS language and the SmartDeployer tool. Then, in Section $3$ we present the Email Processing Pipeline case study, mathematical calculation of system properties like MCL, and we introduce the novel timed architectural modeling/execution language based on our Timed SmartDeployer.
In section $4$, we present our global scaling algorithm and its mathematical foundations.
Finally, in Section $5$ we present simulation of our case study, discussing comparison results, and in Section $6$ we conclude the paper and discuss related work.

\section{Preliminaries}

In this section we present the microservice model, as formalized in \cite{fase_paper,chapter}, the ABS language \cite{abs_docs} and the SmartDeployer tool \cite{fase_paper,chapter}.

%

\subsection{The Microservice Model}\label{mmodel}
%
%
The work in \cite{fase_paper,chapter} formalizes component-based software systems (where 
components are deployed on VMs) 
and the automated deployment problem: synthesis of deployment orchestrations that reach a given target system configuration. 
In particular, the deployment life-cycle of each component type is formalized by means of a finite-state automaton, whose states denote a deployment stage. Each state is associated with a set of provided ports (operations exposed by the component that can be used by other components) and a set of required ports (operations of other components needed for the component to work in that deployment stage).
More specifically, \cite{fase_paper,chapter}
consider the case of 
microservices: components whose deployment life cycle consists of just two phases: $(i)$ creation, which entails \textit{mandatorily} establishing initial connections, via so-called {\it strongly required ports}, with already available microservices, and $(ii)$ subsequent \textit{optional} binding/unbinding, via so-called {\it weakly required ports}, with other microservices. The two phases make it possible to manage circular dependencies among microservices.
These concepts are inspired by Docker Compose~\cite{docker_compose}, a language for defining multi-container Docker applications, that makes it possible for users to specify different relationships among microservices using, e.g.\  the {\sf depends\_on} (resp.\ {\sf external\_links}) modalities that impose (resp.\ do not impose) a specific startup order, in the same way as strong (resp.\ weak) dependencies. 

In addition \cite{fase_paper,chapter} consider resource/cost-aware deployments modeling the memory and computational resources: number of virtual CPU cores (vCores in Azure), sometimes simply called virtual CPUs as in Amazon EC2 and Kubernetes \cite{Hightower}. In particular, both microservice specifications and VM descriptions are enriched with the amount of resources they, respectively, need and supply. 

A microservice {\it deployment orchestration} is a program in an {\it orchestration language} that includes primitives for $(i)$
creating/removing a certain microservice together with its strongly required bindings and $(ii)$ adding/removing weak-required bindings between some created microservices.
Given an initial microservice system, a set of available VMs and a new target set of microservices to be deployed, the \textit{optimal deployment problem} is the problem of finding the deployment orchestration that: satisfies core and memory requirements, leads to a new system configuration including target microservices and optimizes resource usage in case of multiple solutions.  

Differently from the case of components with arbitrary deployment life-cycles 
\cite{sefm-aeolus}, the optimal deployment 
problem has been shown to be decidable for microservices. In particular, 
\cite{fase_paper,chapter}
present a constraint-solving algorithm 
whose result is the new system configuration, i.e.\ the microservices to be deployed, their 
distribution over the VMs and the bindings to be established among 
their strong/weak require and provide ports. 

\subsection{Abstract Behavioral Specification Language}\label{ABS}
Abstract Behavioral Specification  \cite{abs_docs} is an 
actor-based object-oriented specification language (a process algebra) offering algebraic user-defined data types, side effect-free functions and immutable data. 
The ABS toolchain \cite{abs_toolchain} makes it possible to write ABS process algebraic models by conveniently using a programming language syntax and to execute them by means, e.g., of the ABS Erlang backend. 
ABS objects are organized into Concurrent Object Groups (COGs) representing software components or services.
Objects belonging to different COGs communicate with each other using asynchronous method calls \cite{BravettiCZ17}, expressed as \textit{object!method(...)} instructions. Asynchronicity is realized by means of the future mechanism: asynchronous method calls return a future that can be used to wait 
for the result using the \textit{await} statement. 
{\it Timed ABS} is an extension to the ABS core language that introduces a notion of \textit{abstract time}. 
In particular, evolution of time in ABS is modeled by means of discrete time: during execution system time is expressed as the number of {\it time units} that have passed since system start. The modeler decides what a time unit represents for a specific application.
Such a feature makes it possible to perform simulations analysing the time-related behavior of systems. 
Timed ABS has also {\it probabilistic} features that allow modelers to create uniform distributions, e.g.\ the average number of attachments per email in our case study.

To represent 
VMs (and simulate them, e.g., inside the Erlang backend)
ABS introduces the notion of Deployment Component (DC) as a \textit{location} where a COG can be deployed.
As VMs, ABS DCs 
are associated with several kinds of resources. 
In particular virtual cpu speed is represented in ABS by the DC \textit{speed}: it models the amount of {\it computational resource} per time unit a DC can supply to the hosted COGs. This resource is consumed by ABS instructions that are marked with the \textit{Cost} tag, e.g.\ \textit{[Cost: 30] instruction}. COG instructions tagged 
with a cost consume the hosting DC computational resource still available for the current time unit (the instruction above consumes 30 from the DC speed resource): if not enough computational resource is left in the current time unit, then the instruction terminates its execution in the next one.

Concerning the microservice model, in ABS we represent microservice types as classes and instances as objects, each executed in an independent COG. Moreover, we represent strong dependencies as mandatory parameters required by class constructors: such parameters contain the references to the objects corresponding to the microservices providing the strongly required ports. Weak required ports are expressed by means of specific methods that allow an existing object to receive the references to the objects providing them.


\subsection{SmartDeployer}

SmartDeployer implements the algorithm described at the end of Section \ref{mmodel}
to perform automated deployment of 
microservice applications, i.e.\ synthesis of 
deployment orchestrations that reach a given target system configuration.
In particular, it exploits 
the constraint solver Zephyrus2~\cite{zephyrus2}. 
The input to SmartDeployer is expressed by means of
an ABS source file from which it extracts:
\begin{itemize}
\item ABS annotations \textit{[ SmartDeployCost : JSONstring ]}
to {\it classes} representing microservice types. They describe, in JSON format, the functional dependencies (provided and weak/strong required ports) and the resources (number of cores, amount of memory) they need.
\item A global \textit{ [ SmartDeployCloudProvider : JSONstring ]} ABS annotation.
It defines, in JSON format, the types of Deployment Components and their associated resources (e.g.\ number of cores, amount of memory, speed).
\item A global \textit{ [ SmartDeploy : JSONstring ]} ABS annotation. 
It describes, in JSON format, the desired properties of the target configuration, e.g.\ microservice types (possibly with multiple instances) we want to be included in such configuration.
\end{itemize}
In output it produces the 
synthesized {\it deployment orchestration}: the set of {\it orchestration language} instructions (expressed as ABS code) that cause the system to reach a deployment configuration with the desired properties. It also produces the {\it undeployment orchestration} to undo such deployment operations.  A description of the $SmartDeployCost$ annotation can be found in Appendix \ref{SmartDeployCost}.


%

\section{Timed Architectural Modeling/Execution Language}\label{TimeModel}
In this section we introduce our integrated timed architectural modeling/execution language based on the novel {\it Timed SmartDeployer tool}. Our tool fully realizes the integration between timed ABS execution language and architectural annotations by generating {\it timed deployment orchestrations}.  For ease of presentation, we make use of a case study: the Email Pipeline Processing System taken from Iron.io~\cite{ironIO}.
With its help we introduce the concept of microservice Multiplicative Factor (MF) and Maximum Computational Load (MCL).
We show that in our integrated timed language it is possible to model microservice MCL 
in a way that is consistent with timed deployment orchestrations.
As we will see in Section \ref{Adaptation}, this allows us to give a mathematical foundation to the calculation of: the base system configuration and the target ones used by Timed SmartDeployer to synthesize scaling orchestrations (global adaptation algorithm).  
We present the necessary modeling steps and calculations in a conceptual/mathematical way, so that they can be applied to any other microservice application.

\subsection{Case Study and Timed Characteristics of Microservice Systems}
\label{casestudy}
\begin{figure*}[t]
  \centering
  \hspace*{-1.0cm}\includegraphics[scale = 0.33]{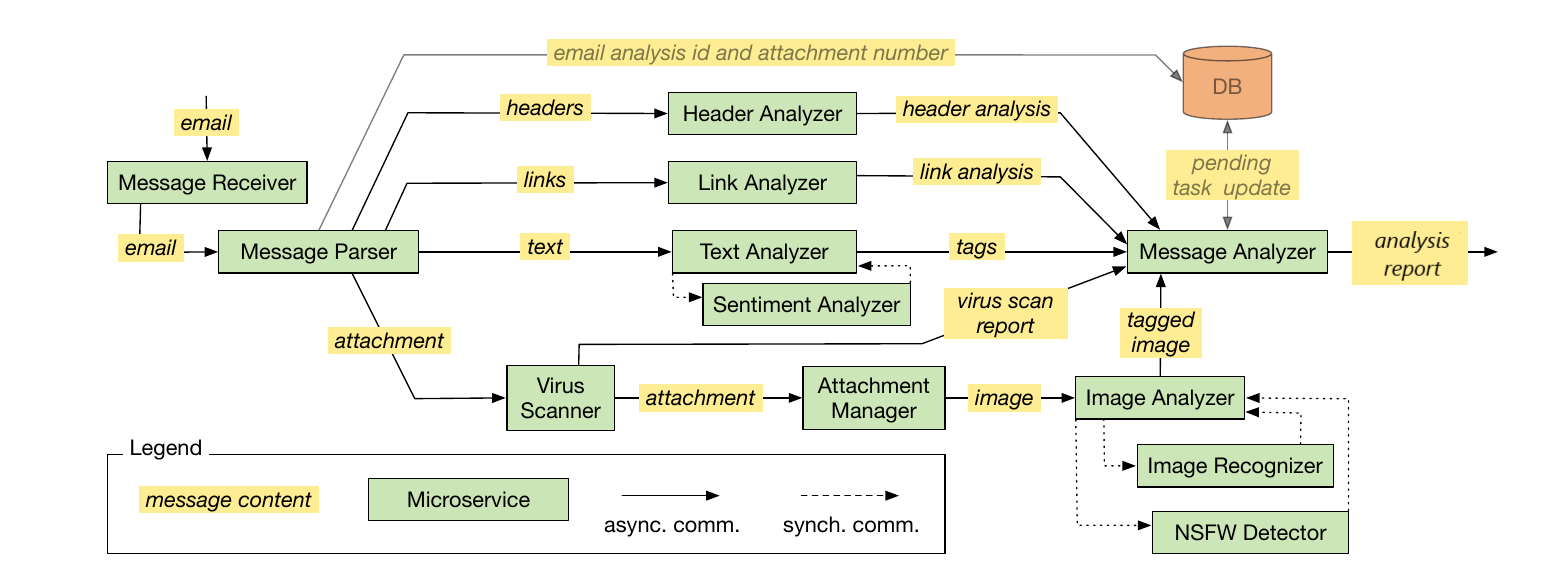}
\vspace*{-.8cm}
  \caption{Microservice Architecture for the Email Processing Pipeline Case Study.}
  \label{fig:arch}
\end{figure*}
In Figure \ref{fig:arch} (similar to that in \cite{fase_paper,chapter})  
we show
the Email Pipeline Processing System 
of~\cite{ironIO}: it is composed of $12$ types of microservices, each one having its own load balancer. The latter is used to distribute requests over a set of instances (connected to weakly required ports) that are incremented/decremented at need.

Recall that in our approach we consider virtual CPU cores, both for machines (providing them) and for microservices (requiring them), see Section \ref{mmodel}. In particular, in our case study, we assume microservices to be deployed on Amazon EC2 VMs of type \textit{large}, \textit{xlarge}, \textit{2xlarge} and \textit{4xlarge}. They respectively provide $2,4,8$ and $16$ virtual CPU cores (following the Azure vCore terminology), simply called \vcpu{}s in Amazon EC2. 
Notice that we model computational resources supplied by VMs (and required by microservices) by means of \textit{virtual} cores with some specified speed, as commonly done by cloud providers 
to abstract underlying hardware.
The cloud provider itself takes care of mapping virtual cores into physical ones by delegating to the runtime (the VM/OS) the scheduling of instructions to make maximal use of real processors. 
Each microservice type is characterized by a \textit{number 
of required virtual cores}. 
Assigning such a number to 
obtain some expected microservice performance (e.g., an expected throughput) is a problem orthogonal to that investigated in this paper. 
While in practice this is usually done as guesswork informed by the experience of the programmers/operators (as in our case), techniques like instruction counting~\cite{BHC06} and profiling~\cite{BHMV09} can help in providing objective estimations of the required cores.

The case study architecture can be divided into four pipelines analyzing different parts of an email. 
Messages enter the system through the \textit{MessageReceiver}, which forwards them to the \textit{MessageParser}. This microservice, in turn, extracts data from the email and routes them to a proper sub-pipeline. Once 
each email component is processed, entailing a specific working time, analysis data is collected by the \textit{MessageAnalyzer} that produces an analysis report.

Based on system architecture,  we observe that each microservice type is also characterized by: $(i)$ a Maximum Computational Load (MCL),  i.e.\ the maximum number of requests that a microservice instance of that type can handle within a second 
and $(ii)$ a Multiplicative Factor (MF) i.e.\ the mean number of requests that a single email entering the system generates for that microservice type.

From a timing viewpoint, considering microservice type MCL and MF is important because 
it allows us to calculate the minimum number of instances of that type needed to guarantee a given overall system MCL $\mathsf{sys\_MCL}$, i.e.\footnote{$\lceil x \rceil$ is the ceil function that takes as input a real number and gives as output the least integer greater than or equal to $x$.}\\[.1cm]
\centerline{$\mathsf{N_{instances} = \ceil[\big]{\frac{sys\_MCL \cdot MF}{MCL}}}$}\\[.1cm]
As we will see in Section \ref{Adaptation}, this is an important system timed characteristic that plays a fundamental role in our global adaptation algorithm. 
\subsection{Microservice MF and MCL Calculation}\label{mclcalc}


The MF of a microservice type is determined from the case study architecture, i.e.\ from the role played by the microservice and 
the email part it receives. As a consequence it
is strictly related to the (average) structure of emails entering the system. In particular we estimate an email to have: 
(i) A single header.
(ii) A set of links (treated collectively as a single information, received by the  \textit{LinkAnalyser}).
(iii) A single text body (received by the  \textit{TextAnalyser}), which is split, on average, into $\mathsf{N_{blocks} =2.5}$ text blocks (individually analysed by \textit{SentimentAnalyser}).
(iv) on
average $\mathsf{N_{attachments} = 2}$ attachments (individually sent to the 
attachment sub-pipeline starting with the 
\textit{VirusScanner}), each having average size of 
$\mathsf{{size_{attachment} = 7MB}}$ and containing a virus with probability $\mathsf{P_V = 0.25}$ (which determines whether a virus scan report is sent to the \textit{MessageAnalyser} or, in case of no virus, the attachment is forwarded to the \textit{AttachmentManager}). 

The average numbers above are estimated ones: the MF of microservices can be easily recomputed in case different numbers are considered. 
In particular, MFs are calculated as follows.
Since emails have a single header, a set of links that are sent together and a single text body, the microservices that analyze these elements, i.e.\ \textit{HeaderAnalyser, LinkAnalyser} and \textit{TextAnalyser},  have $\mathsf{MF = 1}$. As text blocks and attachments are individually sent, each of them generates a request to the \textit{Sentiment Analyser} and the \textit{Virus Scanner}, therefore they have $\mathsf{MF = N_{blocks}}$ and $\mathsf{MF = N_{attachments}}$ respectively. The microservices that follow the \textit{VirusScanner} in the architecture, i.e.\ \textit{AttachmentManager, ImageAnalyzer, ImageRecognizer} and \textit{NSFWDetector} have a MF equal to the number of virus-free attachments, which can be computed as $\mathsf{MF = N_{attachments} \cdot (1 - P_V)}$. Finally, the MF of the \textit{MessageAnalyser} is the sum of the email parts ($1$ header, $1$ set of links, $1$ text body and $\mathsf{N_{attachments}}$ attachments).

The MCL of a microservice is computed as follows:\\[.05cm]
\centerline{$\mathsf{MCL = 1/ (\frac{size_{request}}{data\_rate} + pf)}$}\\[.05cm]
where $\mathsf{size_{request}}$ is the average request size of the microservice in MB. 
Moreover,
$\mathsf{data\_rate}$ is the microservice rate in MB/sec for managing request data. We determine such a value, based on
the number of microservice requested cores, from Nginx server data in \cite{nginx} (considering Nginx servers with
that number of \vcpu{}s).
Finally, $\mathsf{pf}$ is a penalty factor that expresses an additional amount of time that a microservice needs to manage its requests: e.g.\ 
the \textit{ImageRecognizer}, which needs Machine Learning techniques to fulfill its tasks.

We compute microservice $\mathsf{size_{request}}$ as follows. For all microservices receiving attachments, but the \textit{MessageAnalyser} we have:\\[.05cm]
\centerline{$\mathsf{size_{request} = N_{attach\_per\_req} \cdot size_{attachment}}$}\\[.05cm]
where $\mathsf{N_{attach\_per\_req} = N_{attachments}}$ for microservices receiving entire emails and $\mathsf{N_{attach\_per\_req} = 1}$ for the others.
For \textit{HeaderAnalyser, LinkAnalyser} and \textit{TextAnalyser} we consider $\mathsf{size_{request}}$ to be neglectable, thus (since their $\mathsf{pf}$ is also $0$) their $\mathsf{MCL}$ is infinite.
Concerning \textit{MessageAnalyser} request size, we compute the average size of the MF requests that en email entering the system generates (since we consider only attachments to have a non-negligible size), i.e. \\[.05cm]
\centerline{$\mathsf{size_{request\_MA} = \frac{N_{attachments}  \cdot (1 - P_V) \cdot size_{attachment}}{MF}}$.}\vspace{-.25cm}

\subsection{Timed SmartDeployer}\label{timedSD}

Our timed architectural modeling/execution language fully integrates timed ABS and architectural annotations thanks to the novel {\it Timed SmartDeployer}. Such a tool extends SmartDeployer~\cite{fase_paper} with synthesis of {\it timed} deployment orchestrations: they additionally encompass dynamic management of overall Deployment Component (DC) {\it startup time} 
and DC {\it speed} (computational resources per time unit, see Section \ref{ABS}),  based on the number of DC virtual cores that are actually used by some microservice after enacting the synthesized deployment sequence.
As we will show, this allows us to correctly model time (microservice MCL). 

The original SmartDeployer implicitly handles time by simply assigning all properties of DCs, copying them from annotations. The effect of this on timed ABS was to {\it statically} assign
a $\mathsf{speed}$ and a $\mathsf{startup\_time}$ to each DC. 
Concerning $\mathsf{speed}$, this caused microservices, deployed in a DC with unused cores, to unrealistically proceed 
faster:
as if they could exploit the computational power of unused cores. Our solution is to 
dynamically evaluate, during orchestration, the number of DC cores that are actually used by deployed services, and to adjust each DC speed to: $\mathsf{speed}\,\text{-}\,\mathsf{speed\_per\_core} \cdot \mathsf{unused\_cores}$. 
Concerning $\mathsf{startup\_time}$, since in synthesized orchestrations DCs are sequentially created, in timed ABS the overall startup time turned out to be the {\it sum} of that of individual DCs. 
To have a more realistic modeling of virtual machine provisioning (where VMs are contemporaneously acquired), our solution is to dynamically set such a time to the {\it maximum} of their startup time.
The above was realized by automatically synthesizing orchestrations, whose language additionally includes (w.r.t.\ SmartDeployer) 
two primitives {\it explicitly} managing time aspects
\begin{itemize}
\item One to decrement the speed of a DC: \textit{decrementResources(\dots)} in ABS.
\item One to set overall the startup time of created DCs: \textit{duration(\dots)} in ABS.
\end{itemize}

\subsection{Modeling Service MCL}\label{MCLexample}


We now show how Time SmartDeployer allows us to correctly simulate the 
service MCL we want to model (see Section \ref{mclcalc}), independently of the VM (DC) in which it is deployed. 
An example is considering, as we do in our case study, the ABS time unit to be $1/30$ $sec$ and setting  VMs to supply $5$ $\mathsf{speed\_per\_core}$. 
In the ABS code of a service we implement its MCL by using the \textit{Cost} instruction tag (see Section \ref{ABS}). 
E.g., for the 
\textit{ImageRecognizer}, which requires $6$ cores to be deployed, we obtain the 
MCL of $91$ requests per second as follows:
\begin{lstlisting}
class ImageRecognizer() implements ImageRecognizerInterface {
  Int mcl = 91;
  String recognizeImage(String image, ImageRecognizer_LoadBalancerInterface balancer){
    [Cost: 5 * 6 * 30 / mcl] balancer!removeMessage();
    Int category = random(9);
    return "Category Recognized: " + toString(category);}}
\end{lstlisting}
\vspace{-.2cm}
where the method $recognizeImage(...)$ is executed at each request. 

Due to our SmartDeployer timed extension, the amount 
of VM speed used by \textit{ImageRecognizer} is always $\mathsf{5  \cdot 6}$ ($\mathsf{speed\_per\_core \cdot cores\_required}$), independently of the VM in which it is deployed: i.e.\
\textit{ImageRecognizer} can use up to  $\mathsf{5  \cdot 6}$ computational resources per time unit. 
The $Cost$ tag above causes each request to 
consume $\mathsf{speed\_per\_core \cdot cores\_required \cdot 30/MCL}$ computational resources. 
Therefore, since $\mathsf{MCL/30}$ is the \textit{ImageRecognizer} MCL expressed in requests per time unit, this realizes the desired (deployment independent) service MCL.

%
%
%
%
%

\section{Global Run-Time Adaptation}\label{Adaptation}

In this section, we present our algorithm for global run-time adaptation, which is totally independent from the case study (and from the ABS language itself).

\subsection{Calculation of Scaling Configurations}\label{scalecalc}

We consider a base \textbf{B} system configuration, see Table \ref{table:tab2}, which guarantees a system MCL of $60$ emails/sec.
In the corresponding column of Table \ref{table:tab2} we present the number of instances for each microservice type, calculated according to the formula in Section \ref{casestudy}.
Moreover, we consider four incremental configurations $\Delta 1$,
$\Delta 2$, $\Delta 3$ and $\Delta 4$, synthesized via Timed SmartDeployer, each adding a number of instances to each microservice type, see Table \ref{table:tab2}.
Those incremental configurations are used as target configurations for deployment/undeployment orchestration synthesis
in order to perform run-time architecture-level reconfiguration.
As shown in Table \ref{table:tab3}, $\Delta 1$, $\Delta 2$, $\Delta 3$ and $\Delta 4$  are used, in turn, to build (summing up them element-wise as arrays) the incremental configurations $\mathsf{Scale1, Scale2, Scale3}$ and $\mathsf{Scale4}$ that guarantee an additional system MCL of $+60$, $+150$, $+240$ and $+330$ emails/sec, respectively.

The reason for not considering our $\mathsf{Scales}$ as monolithic blocks and defining them as combinations of the $\Delta$ incremental configurations is the following. 
Let us suppose the system to be, e.g., in a \textbf{B} $\! + \!$ $\mathsf{Scale1}$ configuration and the increase in incoming workload to require the deployment of $\mathsf{Scale2}$ and the undeployment of $\mathsf{Scale1}$. If we had not introduced $\Delta$ configurations and we had synthesized orchestrations directly for $\mathsf{Scale}$ configurations, we would have needed to perform an undeployment of $\mathsf{Scale1}$ followed by a deployment of $\mathsf{Scale2}$. With $\Delta$ configurations, instead, we can simply additionally deploy $\Delta 2$.  Moreover, notice that dealing with such an incoming workload increase by naively deploying another $\mathsf{Scale1}$ additional configuration, besides the already deployed one, would not lead the 
system MCL to be increased of another $+60$ emails/sec. This is because the maximum number of email per seconds that can be
handled by individual microservices composing the obtained $\mathsf{\textbf{B} \! + \! 2 \! \cdot \! Scale1}$ configuration would be unbalanced.
Such an effect worsens if the system incoming workload keeps slowly increasing and further additional $\mathsf{Scale1}$ configurations are deployed. Since $\mathsf{Scale1}$ for some microservices (\textit{AttachmentManager, ImageAnalyser}) does not provide additional instances, such microservices would eventually become the bottleneck of the system and the system MCL would no longer increase. 
Moreover, 
$\Delta$ configurations yield, w.r.t.\ monolithic $\mathsf{Scale}$ ones, 
a finer granularity
that makes SmartDeployer orchestration synthesis faster. 

For each microservice type, the number of additional instances considered in Tables \ref{table:tab2} and \ref{table:tab3} for the $\mathsf{Scale}$ configurations have been calculated as follows. Given the additional system MCL to be guaranteed, the number $\mathsf{N_{deployed}}$ of instances of that microservice already deployed and its MF and MCL, we have:\\[.1cm]
\centerline{$\mathsf{N_{instances} = \ceil[\big]{\frac{(base\_MCL + additional\_MCL) \cdot MF}{MCL} - N_{deployed}}}$}

In the following section we will present the algorithm for global adaptation. 
The algorithm is based on the principles described here, i.e.\ it has the following {\it invariant} property: if $\mathsf{N}$ $\mathsf{Scale}$ configurations are considered ($\mathsf{N=4}$ in our case study) and are indexed in increasing order of additional system MCL they guarantee, the system configuration reached after adapting to the monitored inbound workload is either 
\textbf{B} or $\mathsf{\textbf{B} + (n \cdot ScaleN) + scale}$, for some $\mathsf{scale \in \{Scale1, Scale2, \dots, ScaleN\}}$ and $\mathsf{n \geq 0}$. 
The invariant property indeed shows, as we explained above, that the deployment of sequences of the same $\mathsf{Scale}$ configuration is not allowed, except for sequences of $\mathsf{ScaleN}$. 
This is because, the biggest configuration $\mathsf{ScaleN}$ should be devised, for the system being monitored, in such a way that the inbound workload rarely yields to additional scaling needs. Moreover, even if a sequence of $\mathsf{ScaleN}$ occurs, the system would be sufficiently balanced. This is because, differently from smaller $\mathsf{Scale}$ configurations, $\mathsf{ScaleN}$ is assumed to add, at least, an instance for each microservice having non-infinite MCL (as for $\mathsf{Scale4}$ in our case study). 

\begin{table}[t]
\scalebox{1}{
\begin{tabular}{ |c|c|c|c|c|c|c|c|c|c|c|c| } 
 \hline
 \textbf{Microservice} & \textbf{B} & $\Delta1$ & $\Delta2$  & $\Delta3$ & $\Delta4$ & \textbf{Microservice} & \textbf{B} & $\Delta1$ & $\Delta2$  & $\Delta3$ & $\Delta4$  \\
 \hline
 \textbf{Message Receiver}  & 1 & +1 & +0 & +1 & +1 & \textbf{Virus Scanner} & 1 & +1 & +2 & +1 & +2 \\
 \hline
 \textbf{Message Parser} & 1 & +1 & +0 & +1 & +1 & \textbf{Attachment Manager} & 1 & +0 & +1 & +0 & +1 \\
 \hline
 \textbf{Header Analyser} & 1 & +0 & +0 & +0 & +0 & \textbf{Image Analyser} & 1 & +0 & +1 & +0 & +1 \\
 \hline
 \textbf{Link Analyser} & 1 & +0 & +0 & +0 & +0 & \textbf{NSFW Detector} & 1 & +1 & +2 & +1 & +2   \\
 \hline
 \textbf{Text Analyser} & 1 & +0 & +0 & +0 & +0 &  \textbf{Image Recognizer} & 1 & +1 & +2 & +1 & +2 \\
 \hline
 \textbf{Sentiment Analyser} & 2 & +1 & +3 & +2 & +2 & \textbf{Message Analyser} & 1 & +1 & +2 & +1 & +2  \\
 \hline
\end{tabular}
}
\vspace*{.1cm}
\caption{Base \textbf{B} (60 $\frac{emails}{sec}$) and incremental $\Delta$ configurations.}
\label{table:tab2}
\end{table}

\begin{table}[t]
\vspace*{-.2cm}
\scalebox{0.88}{
\hspace*{-0.2cm}\begin{tabular}{ |c|c|c|c| } 
 \hline
 \textbf{Scale 1 (+60 $\frac{emails}{sec}$)} & \textbf{Scale 2 (+150 $\frac{emails}{sec}$)} & \textbf{Scale 3 (+240 $\frac{emails}{sec}$)} & \textbf{Scale 4 (+330 $\frac{emails}{sec}$)} \\
 \hline
 $\Delta1$  & $\Delta1 + \Delta2$  & $\Delta1 + \Delta2 + \Delta3$ & $\Delta1 + \Delta2 + \Delta3 + \Delta4$   \\
 \hline
\end{tabular}
}
\caption{Incremental $\mathsf{Scale}$ configurations.}
\label{table:tab3}
\end{table}


\subsection{Scaling Algorithms}\label{scalealg}

For comparison purposes, we realized two algorithms,  for local and global adaptation. 
In both of them we use a scaling condition on monitored inbound workload 
involving two constants called $\mathsf{K}$ and $\mathsf{k}$. $\mathsf{K}$ is used to leave a margin under 
the guaranteed MCL, so to make sure that the system can handle the 
inbound workload. $\mathsf{k}$ is used to prevent fluctuations, i.e.\ sequences 
of scale up and down.

The condition for scaling up is $\mathsf{(inbound\_workload  + K) - total\_MCL  > k}$ and the one for scaling down is $\mathsf{total\_MCL - (inbound\_workload + K) > k}$.
The interpretation of such conditions changes, depending on whether they are used for the local or global adaptation algorithm. In the case of local adaptation the conditions are applied by monitoring a single microservice type: $\mathsf{inbound\_workload}$ is the number of requests per second received by the microservice load balancer and $\mathsf{total\_MCL}$ is the $\mathsf{MCL}$ of a microservice instance of that type (calculated as explained in Section \ref{mclcalc}) multiplied by the number of deployed instances. In the case of global adaptation the conditions are applied by monitoring the whole system: $\mathsf{inbound\_workload}$ is the number of requests (emails in our case study) per second entering the system and $\mathsf{total\_MCL}$ is the system $\mathsf{MCL}$.
A detailed explaination of the local adaptation algorithm can be found in Appendix \ref{localAdaptation}.

Concerning global adaptation, we have a single monitor that periodically executes (e.g.\ every 10 seconds in our case study)
the code excerpt below. The code uses constants $\mathsf{numScales}$, representing the number of $\mathsf{Scale}$ configurations ($4$ in our case study), and $\mathsf{scaleComponents}$: an array\footnote{The ABS instructions $\mathsf{nth(a,i)}$ and $\mathsf{length(a)}$ retrieve the $\mathsf{i}$-th element and the length of the $\mathsf{a}$ array, respectively.} of $\mathsf{numScales}$ elements (corresponding to Table \ref{table:tab3} in our case study) that stores in each position an array representing a $\mathsf{Scale}$ configuration (i.e.\ specifying, for each microservice, the number of additional instances to be deployed).
Moreover, the code uses the variables $\mathsf{sys\_MCL}$, containing the current system MCL
(assumed to be initially set to the \textbf{B} configuration MCL,  see Table \ref{table:tab2} in our case study), and $\mathsf{deployedDeltas}$:
an array of $\mathsf{numScales}$ numbers that keeps track of the number of currently deployed $\Delta$ incremental configurations
(assumed to be initially empty, i.e.\ with all 0 values). Both variables are updated by the code in case of scaling.
First of all the code applies the above described scale up/down conditions. Then it loops, starting
from the \textbf{B} configuration in variable $\mathsf{config}$ (an array that stores, for each microservice, the number of instances we currently consider), and selecting $\mathsf{Scale}$ configurations to add to $\mathsf{config}$, until a configuration $\mathsf{c}$ is found such that its system MCL 
satisfies $\mathsf{sys\_MCL - (inbound\_workload + K) \geq 0}$. The system MCL of a configuration $\mathsf{c}$ is calculated with method \textit{mcl}, which yields \\[.05cm]
\centerline{$\mathsf{min_{1 \leq i \leq length(config)} \; \; nth(config,i \!-\! 1) \cdot MCL_i / MF_i}$}\\[.1cm]
with $\mathsf{MCL_i/MF_i}$ denoting the MCL/MF of the $\mathsf{i}$-th microservice. 
More precisely the algorithm uses an external loop updating variables $\mathsf{config}$ and $\mathsf{configDeltas}$ according to the 
incremental $\mathsf{Scale}$ selected by the internal loop: $\mathsf{configDeltas}$ is an array with
the same structure of $\mathsf{deployedDeltas}$, which is initially empty and, every time a $\mathsf{Scale}$ configuration is selected, is updated by incrementing the amount of corresponding $\Delta$ configurations (as described in Table \ref{table:tab3} in our case study).
The internal loop selects a $\mathsf{Scale}$ configuration by looking for the first one that, added to $\mathsf{config}$, yields a candidate configuration whose system MCL satisfies the condition above. If such $\mathsf{Scale}$ configuration is not found then it just selects the last (the biggest) $\mathsf{Scale}$ configuration ($\mathsf{Scale4}$ in our case study), thus implementing the invariant presented in Section \ref{scalecalc}.




\begin{lstlisting} 
if((inbound_workload+kbig)-sys_MCL>k || (sys_MCL-(inbound_workload+kbig)>k){
 List<Int> configDeltas = this.createEmpty(numScales);
 List<Int> config = baseConfig;
 sys_MCL = this.mcl(config);
 Bool configFound = sys_MCL-(inbound_workload+kbig)>=0;
 while(!configFound) {
  List<Int> candidateConfig = baseConfig;
  Int i = -1;
  while(i<numScales-1 && !configFound){
    i=i+1;
    candidateConfig = this.vectorSum(config,nth(scaleComponents,i));
    sys_MCL = this.mcl(candidateConfig);
    configFound = sys_MCL-(inbound_workload+kbig)>=0;}
  config = candidateConfig;
  configDeltas = this.addDeltas(i,configDeltas);}
 this.reconfigureSystem(deployedDeltas,configDeltas);
 deployedDeltas = configDeltas;}
\end{lstlisting}

Finally, as we show in the method \textit{reconfigureSystem} below, given the target $\Delta$ configurations $\mathsf{configDeltas}$ to be reached and the current $\mathsf{deployedDeltas}$ ones, we perform the 
difference between them so to find the $\Delta$ orchestrations that have to be (un)deployed.
\begin{lstlisting}
Unit reconfigureSystem(List<Int> deployedDeltas, List<Int> configDeltas) {
  Int i = 0;
  while(i<numScales) {
    Int diff = nth(configDeltas,i)-nth(deployedDeltas,i);
    Rat num = abs(diff);
    while(num>0) {
      if (diff>0) {nth(orchestrationDeltas,i)!deploy();}
      else {nth(orchestrationDeltas,i)!undeploy();}
      num = num-1;}
    i = i+1;}}
\end{lstlisting}
We use methods \textit{deploy/undeploy} of the object in the position $i \! - \! 1$ of the array $\mathsf{orchestrationDeltas}$ to execute the orchestration of the $\mathsf{i}$-th $\Delta$ configuration. 
In our model such an orchestration is the ABS code generated by Timed SmartDeployer at compile-time: it makes use of ABS primitives \textit{duration(\dots)} and \textit{decrementResources(\dots)} to \textit{dynamically} set, respectively, the overall startup time to the maximum of those of deployed DCs and the speed of such DCs accounting for the virtual cores actually being used (by decrementing the DC static speed, see Section \ref{timedSD}). In this way we are guaranteed that each microservice always preserves the desired fixed MCL we want to model (see Section \ref{MCLexample}). Moreover, we remind that, besides speed,  also constraints related to other resources (memory) are considered in the SmartDeployer synthesis process. 


\section{Simulation with ABS}

In this section we present simulation results obtained with our ABS programs~\cite{ABS_simulations} modeling local and global scaling (via Timed SmartDeployer orchestrations) for our 
case study. Such programs 
encompass, besides static aspects of the case study 
architecture (annotations), 
also 
the code representing 
service/adaptation behavior {\it under an inboud workload}: 
they fully implement what we explained in Sections 
\ref{TimeModel} and \ref{Adaptation}.
In particular, we implement by means of \textit{monitoring services}: our algorithm for global adaptation (a single system monitor) and the one for local adaptation (a monitor for each load balancer) by just detecting scaling needs and enacting replications at the level of single microservices. Monitors are implemented by dedicated ABS services that 
run on a separate (simulated) VM. For these services we do not model the computing resources:
we assume that monitors
are part of the deployment infrastructure, which is also
responsible for enacting the scaling strategies
(as it happens, e.g, with Kubernetes autoscaling).


To make scaling operations realistic, it is important to explicitly represent VM overall startup time and, within load balancers, request queues of a fixed size. This explicit management not only provides a realistic model, but is also crucial for preventing the system from over-loading. Indeed, without these queues, the system wouldn't refuse any message and when the inbound workload grows up, it would overload the system with no possibility of restoring acceptable performances even if scaling actions occur. Moreover, queues allow us to model message loss and to use it for comparing the behavior of local and global scaling. 
In our modeling, we assume microservices not to fail and messages to be eventually delivered unless the receiver queue is overloaded (in this case they are dropped).  

We decided to test our approach using both a real 
diurnal load pattern inspired to that in \cite{realWorkload}, see Figure \ref{fig:mess}, and part of an IMAPS email traffic similar to that in \cite{KaramollahiW19} (accounting for the fact that here email attachments are also considered), 
see Figure \ref{fig:mess2}. 
We implemented such inbound workloads by means of an {\it email generating service}.  
The ABS code is executed with the Erlang backend. 



\subsection{Simulation Results}


We compare the simulation of our approach based on global scaling with the classical one (based on local scaling) by focusing on the following aspects: (i) latency comparison, (ii) message loss comparison and (iii) number of microservices comparison.
\begin{figure}[t]
\vspace{-.2cm}
\subfloat[Diurnal Load Pattern]{\includegraphics[height=1.2in, width=2.4in]{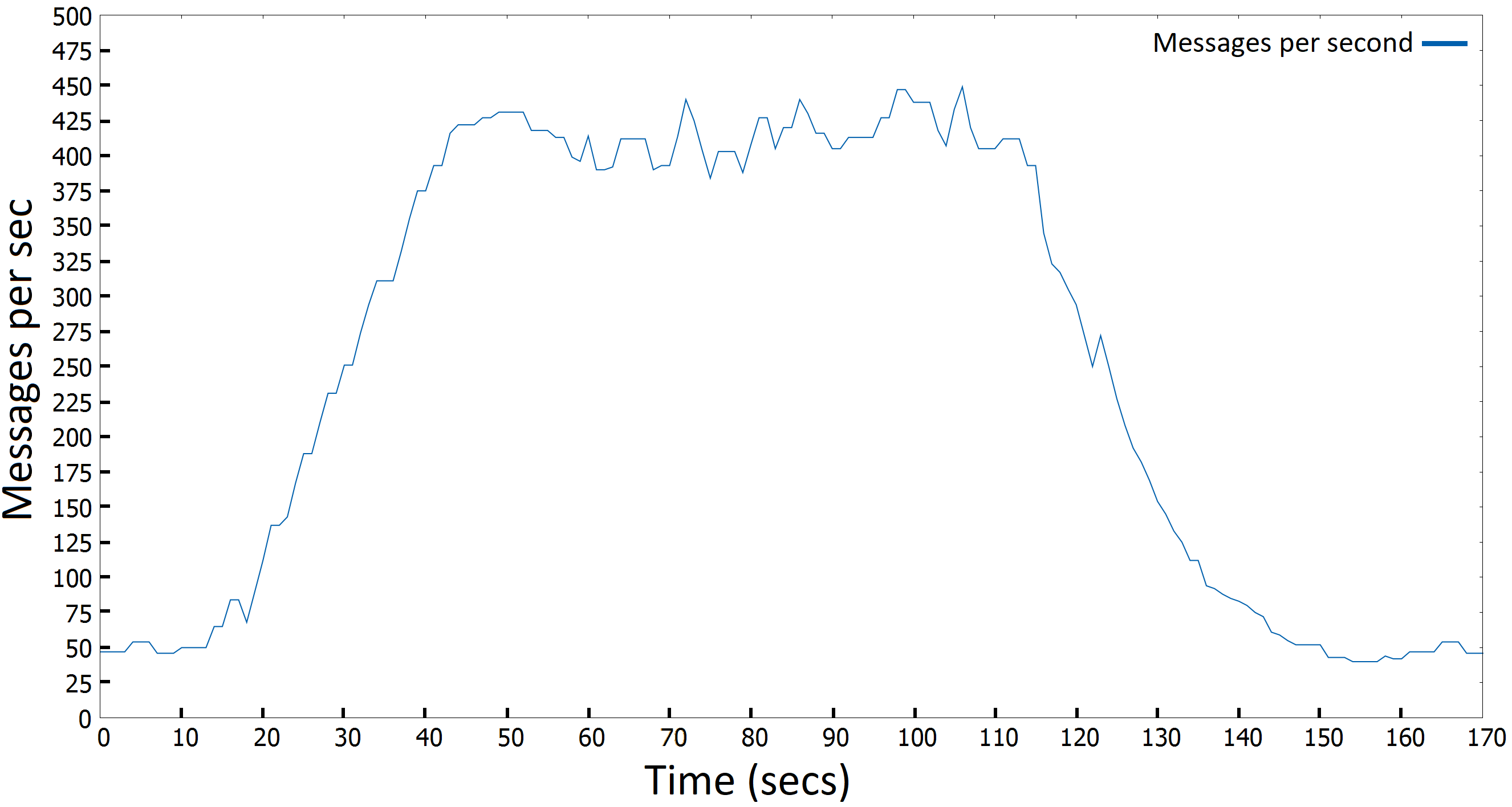} \label{fig:mess}}
\subfloat[Latency Comparison]{\includegraphics[height=1.2in, width=2.4in]{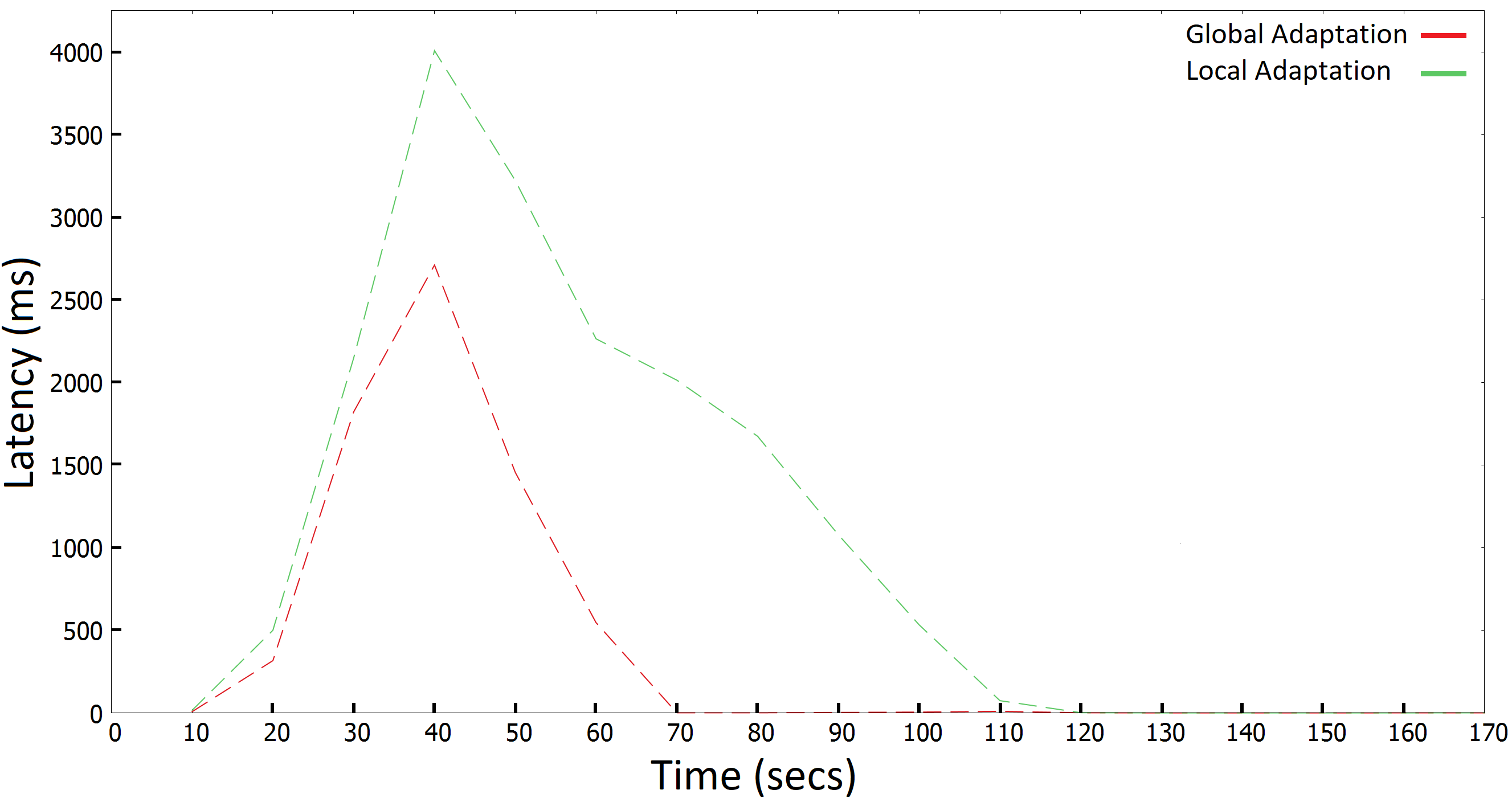} \label{fig:lat}}
 \newline
\subfloat[Loss Comparison]{\includegraphics[height=1.2in, width=2.4in]{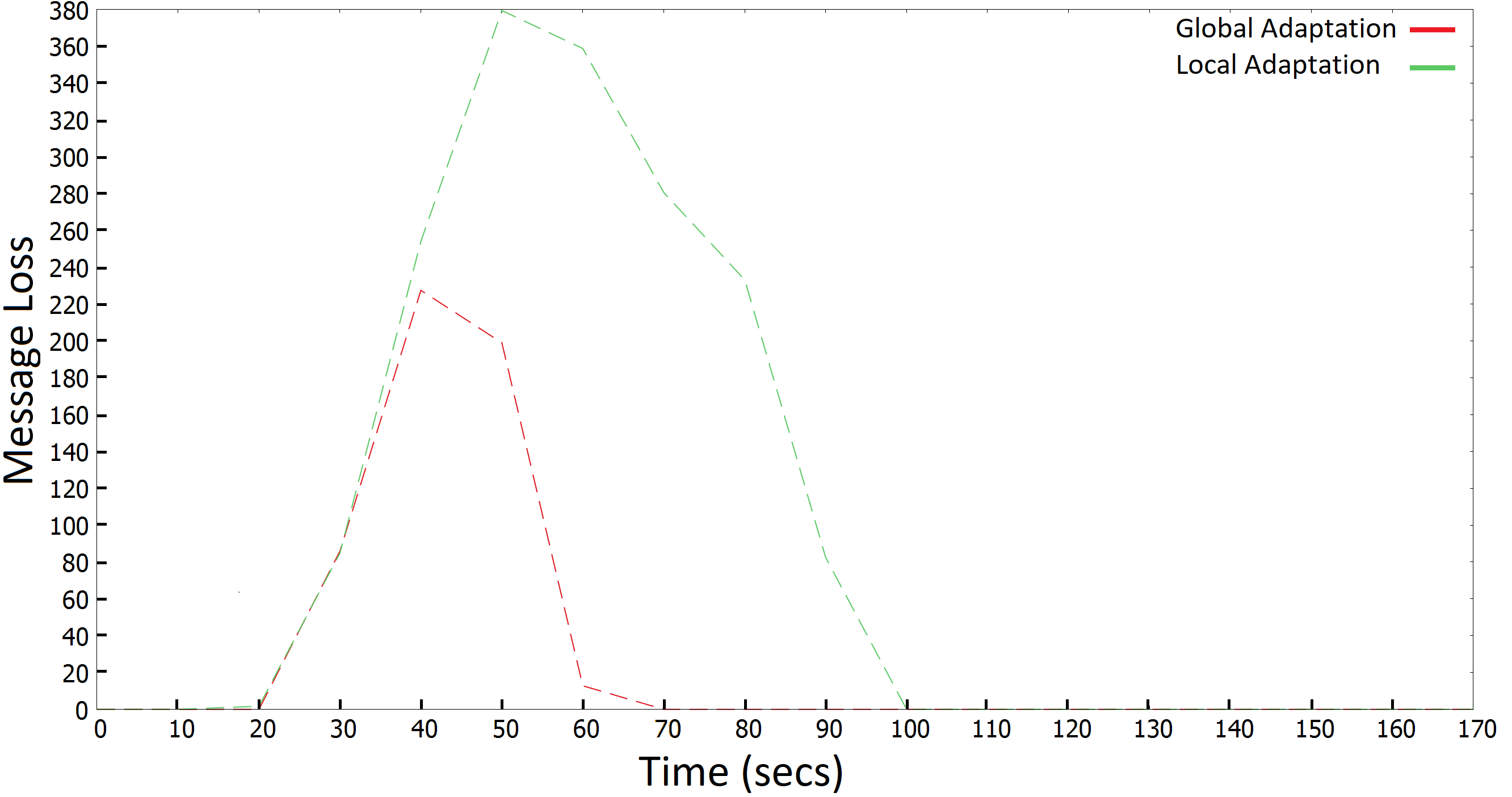} \label{fig:loss}}
\subfloat[Number of Microservices]{\includegraphics[height=1.2in, width=2.4in]{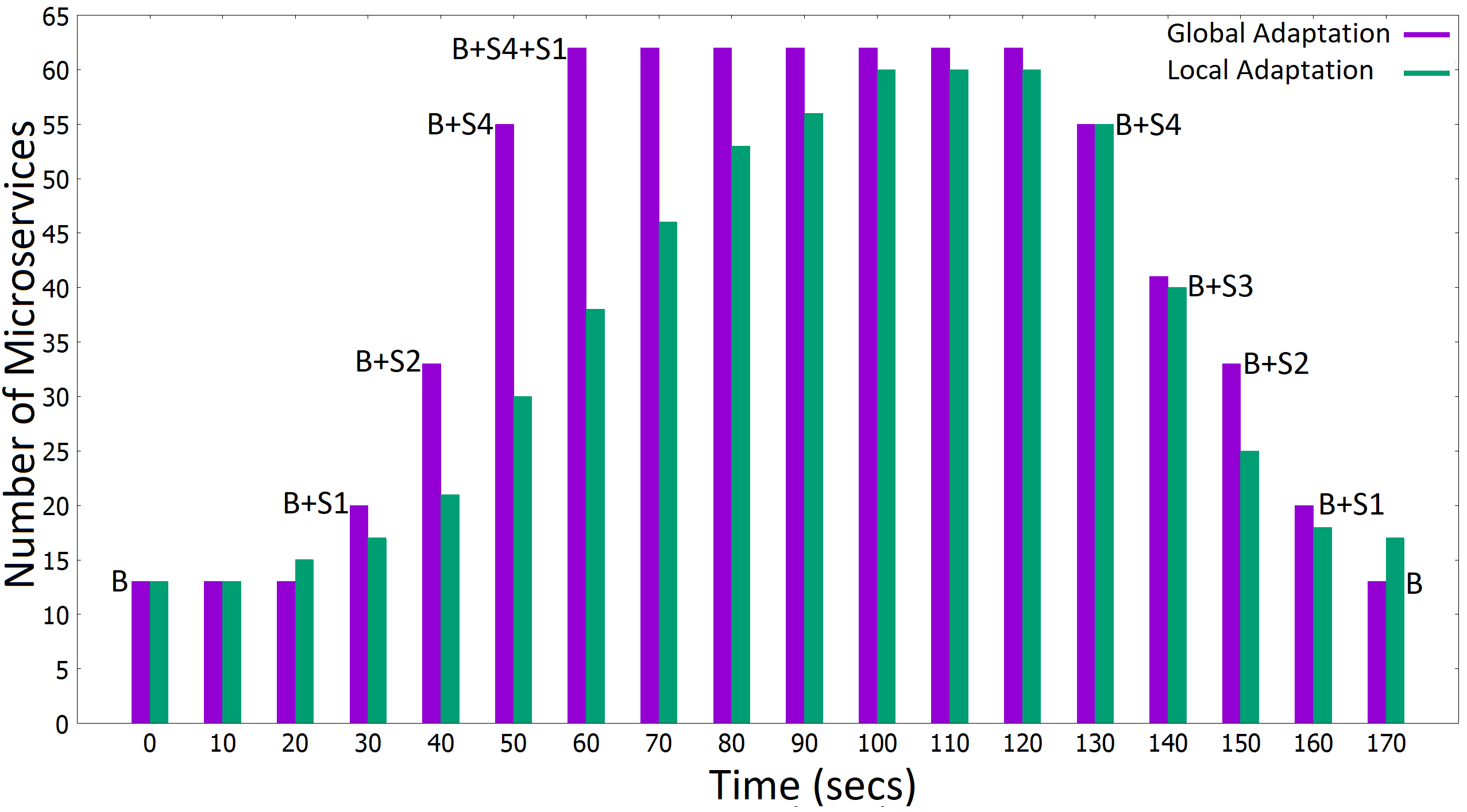} \label{fig:comp}}
 \caption{Comparison results under the real diurnal load pattern.}
  \centering
 \end{figure}
\begin{figure}[t]
\vspace{-.2cm}
\subfloat[IMAPS Email Traffic]{\includegraphics[height=1.2in, width=2.4in]{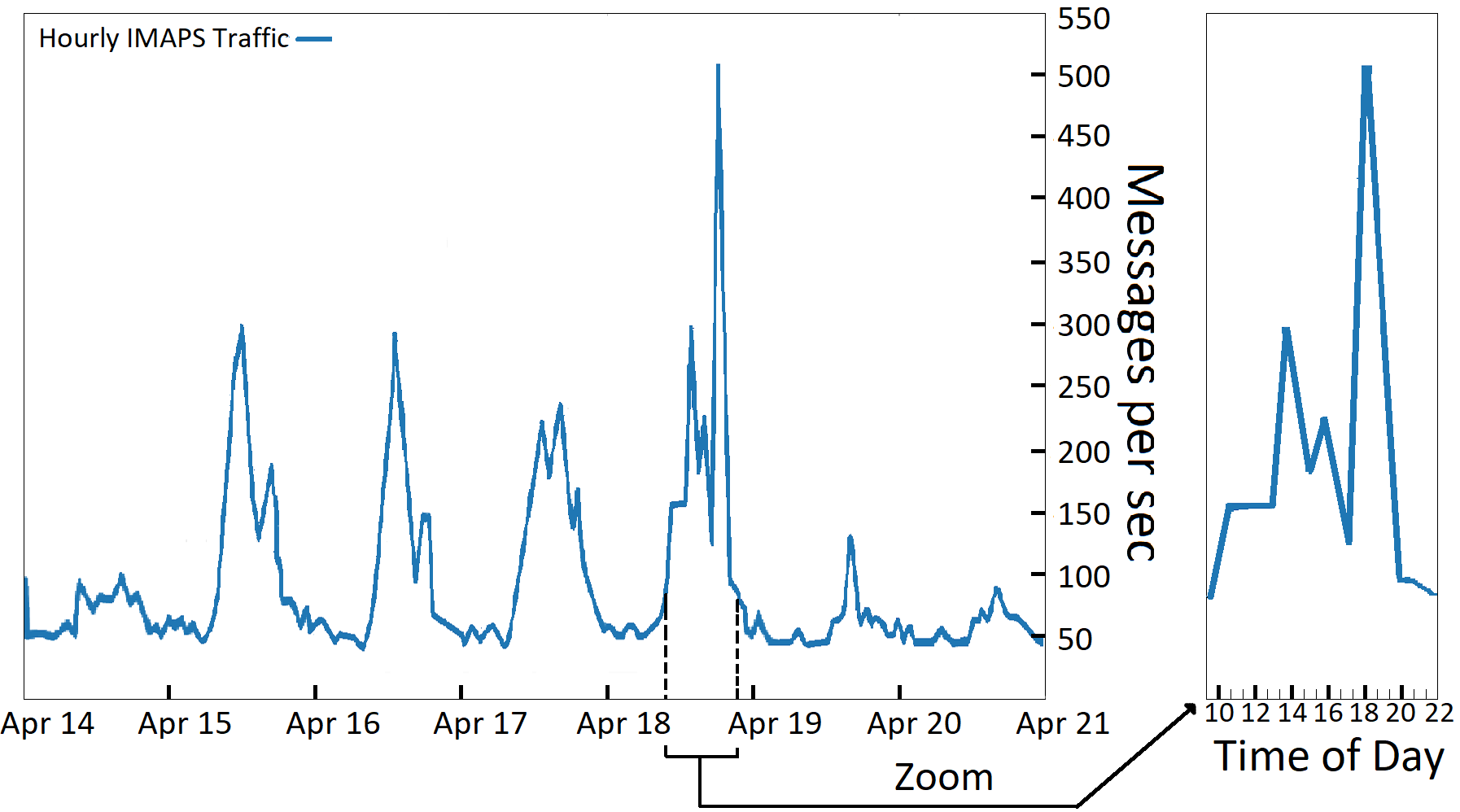} \label{fig:mess2}}
\subfloat[Latency Comparison]{\includegraphics[height=1.2in, width=2.4in]{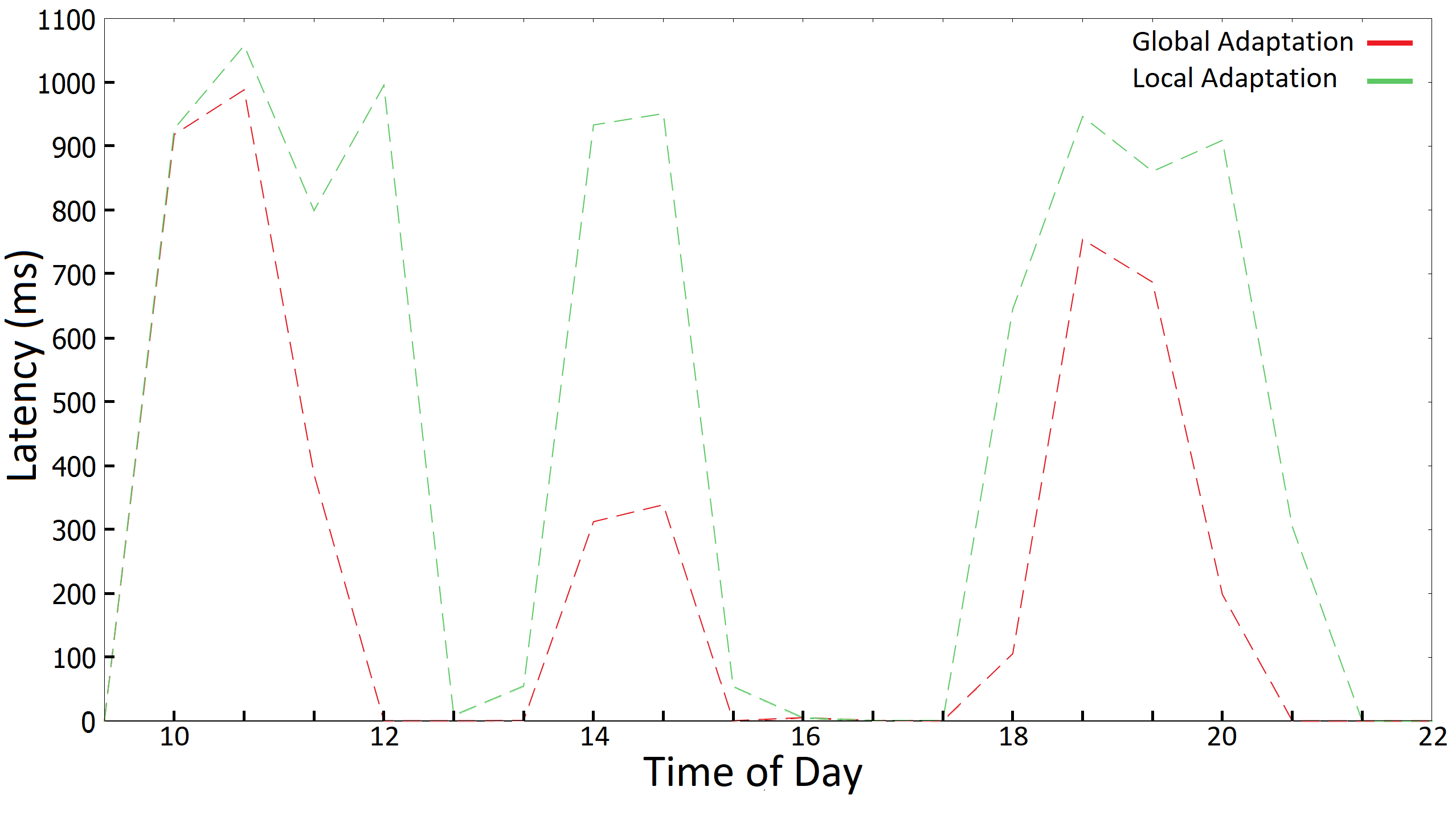} \label{fig:lat2}}
 \newline
\subfloat[Loss Comparison]{\includegraphics[height=1.2in, width=2.4in]{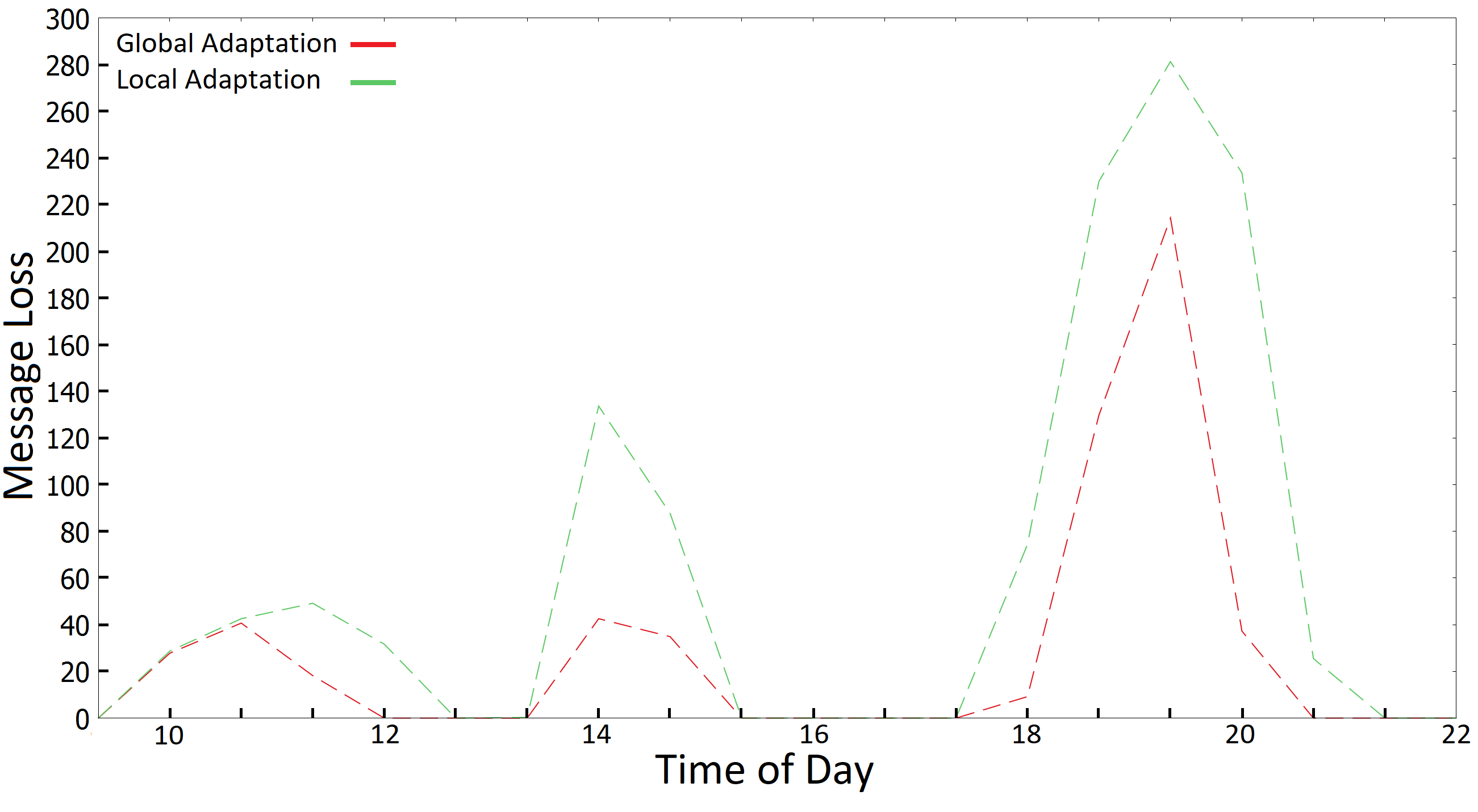} \label{fig:loss2}}
\subfloat[Number of Microservices]{\includegraphics[height=1.2in, width=2.4in]{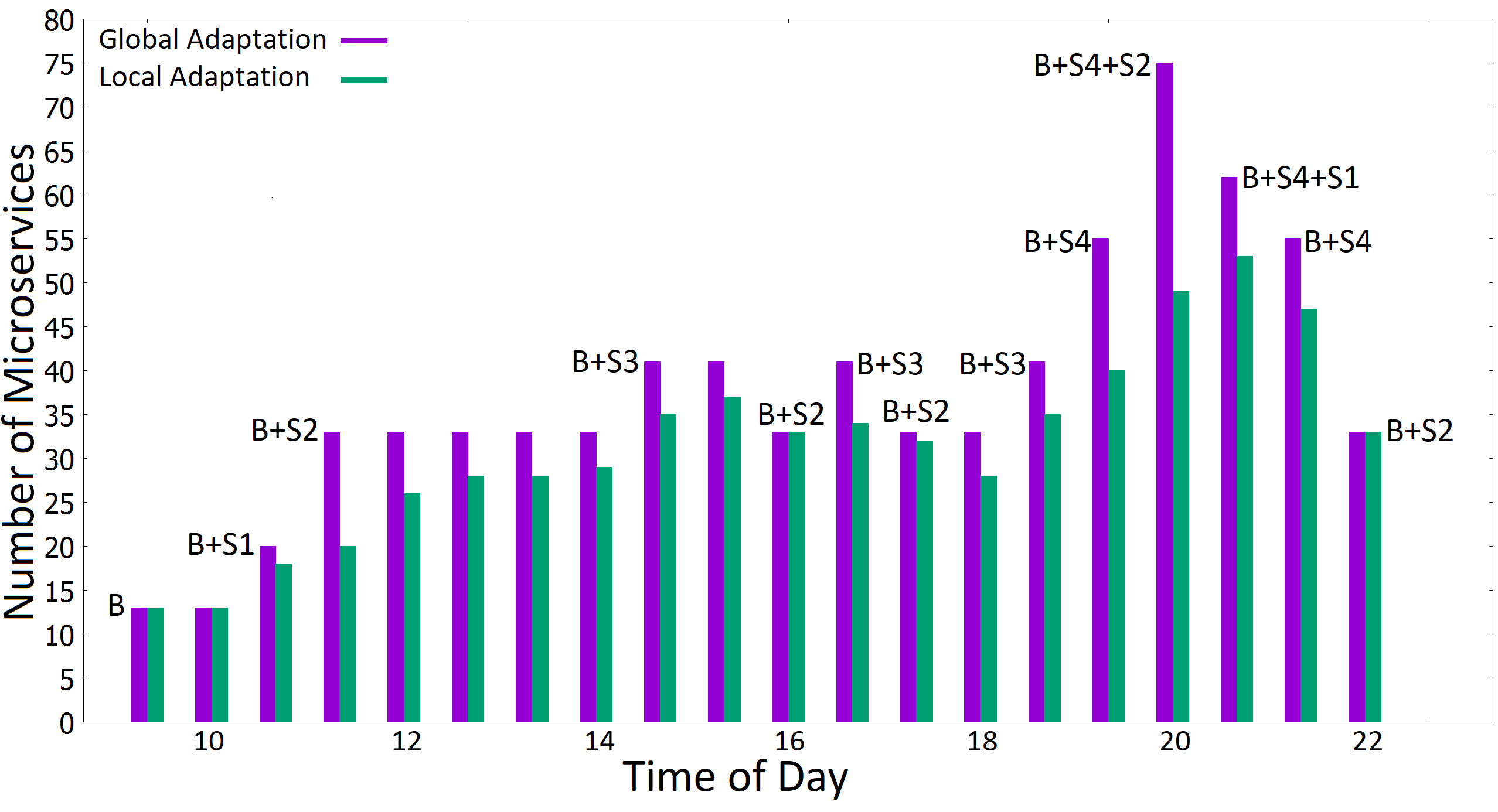} \label{fig:comp2}}
 \caption{Comparison results under the IMAPS email traffic.}
  \centering
\end{figure}
The first metric to be analyzed, in order to evaluate the performance of our new scaling approach, is the {\it latency}. We consider the latency as the average time for completely processing an email that enters the system. As shown by Figures \ref{fig:lat} and \ref{fig:lat2} (the latter considers monitoring time
to be $40$ mins instead of $10$ secs), our approach, represented by the red dashed line, is outperforming the classical one. Considering the different peaks of incoming messages present in the chosen workloads, it is clear the extent of the improvement introduced by our new approach: our global adaptation makes the system adapt much faster than the classic approach. This is caused by the ability of the global adaptation strategy of detecting in advance the scaling needs of all system microservices.

The above observation is confirmed by analyzing system {\it message loss}. Observing Figures \ref{fig:loss} and \ref{fig:loss2}, it is possible to see that our approach always stops losing messages earlier than the classic approach. This means that message queues start to empty and latency can start to decrease.

Finally, comparing the {\it number of deployed microservices} helps to have a deeper understanding of the reasons why the global adaptation performs better. 
As shown by Figures \ref{fig:comp} and \ref{fig:comp2} (where we also label the diagram with the structure of configurations in the case of global scaling), our approach reaches the target configuation, needed to handle the maximum inbound workload, faster than the classical approach. As expected this increments the adaptation responsiveness to higher workloads. The local adaptation slowness in reaching such a target configuration is caused by a \textit{scaling chain effect}: local monitors periodically check the workload, thus single services scale one after the other. Hence, w.r.t.\ global adaptation, in which microservices in the target configuration are deployed together, the number of instances grows slower.
For example, considering
the attachment pipeline in Figure \ref{fig:arch}, the first microservice to 
become a bottleneck is the \textit{VirusScanner}: it starts losing messages, 
which will never arrive to the \textit{AttachmentManager}. Therefore, this component will 
not perceive the increment in the inbound emails until the \textit{VirusScanner} will be replicated, thus causing a scaling chain effect that delays 
adaptation. This is the main cause for the large deterioration in performances 
observed.  
On the other hand, the local approach requires, in total, less resources: this is particularly visible in Figure~\ref{fig:comp}. Due, however, to optimal resource allocation of SmartDeployer reconfigurations, this does not necessarily 
imply a significant
increase in VM costs.

\section{Related Work and Conclusion}

We introduced an integrated timed architectural modeling/execution language that correctly deals with service Maximum Computational Load (MCL). Moreover, we proposed a novel global scaling algorithm that optimally chooses deployment orchestrations, so to keep the system in a configuration that better fits the inbound workload (with the minimum number of instances). Finally, we performed a comparison between our global scaling algorithm and a classical local one by simulating, under two real workloads, a microservice application. 

We now discuss related literature by first comparing with our previous work. In \cite{fase_paper,chapter} initial ideas about applying SmartDeployer generated orchestrations to the case study of \cite{ironIO} were discussed, but (apart from annotations modeling static aspects of the architecture) no actual ABS code implementing system service execution/scaling mechanism
was presented. Moreover, \cite{fase_paper,chapter} draft some scaling configurations just for exemplifying the idea of global adaptation via deployment orchestrations (without presenting any actual scaling algorithm). Such manually drafted scaling configurations are completely different from those here presented in 
Section \ref{scalecalc}, which are precisely calculated (based on service MCL) via a formula yielding the additional number of instances. 
As explained in Section~\ref{scalecalc}, the novel idea of relying on service MCL (and to its mathematical evaluation, see Section \ref{mclcalc}) 
makes it possible to effectively use such configurations in the context of a global scaling algorithm that is guaranteed to reach any target system MCL.
%
%
%
%
%
%
Finally, here we introduce the novel non-monolithic $\Delta$ scales and provide the implementation of the global scaling algorithm. Such algorithm avoids bottlenecks by keeping the system balanced (w.r.t.\ microservice instance number), thanks to the ability of the novel Timed SmartDeployer of correctly dealing with service MCL, see Section \ref{MCLexample}.

We then consider 
additional related work on SmartDeployer. 
While \cite{DEGOUW2019108} just exemplifies the execution of deployment orchestrations for a specific system reconfiguration and \cite{BezirgiannisBG17} additionally deals with selection among different scaling actions based on human suggestions,
we devise: a general methodology for designing a set of deployment orchestrations based on target incremental system MCLs (hence having a mathematical foundation) and an auto-scaling algorithm that makes human intervention unneeded.
Moreover, w.r.t.\  \cite{BezirgiannisBG17,DEGOUW2019108}, we correctly model real aspects such as deployment time and MCL-preserving core-based 
VM speed computation (thanks to our Timed SmartDepoyer) and we also test the effectiveness of our  
algorithm, by comparing it with classical local adaptation.

Regarding related work on auto-scaling, there are several solutions \cite{cloud_watch,mesos,swarm,Hightower} supporting the automatic system reconfiguration, by incrementing or decrementing of the number of instances at the service/container level, when some conditions (e.g., CPU average load greater than 80\%) are met. Our work shows how we can go beyond such local horizontal scaling policies (analyzed, e.g., in \cite{DBLP:conf/tgc/BravettiGGT07}). 

A strand of work sees the predictive capabilities of machine learning applied to
auto-scaling. Below, we cite a few relevant examples, but we point the
interested reader to the survey in~\cite{LML14} for a more
comprehensive view on the field. In \cite{HXW20} a
scheduling system is proposed, which is based on deep reinforcement learning. There, the scheduler
interacts with the deployment environment to learn scheduling strategies without
any prior knowledge of both the environment and the services. Similarly, ~\cite{HA18} attacks the problem of defining optimal thresholds
for scaling policies with a reinforcement-learning algorithm that automatically
and dynamically adjusts the thresholds without user configuration. Finally, ~\cite{MLFog} proposes an approach that uses a predictive autoscaling model
trained on a dataset generated from simulations of reactive rule-based
autoscaling.
W.r.t.\ work on workload prediction, such as~\cite{MLFog}, our global adaptation algorithm ability of detecting in advance service scaling needs is not based on guessing workload by means of logged data, but on  mathematically calculating service MCL from system MCL (thanks to service Multiplicative Factor and current number of instances, see formula in Section~\ref{casestudy}).  The two approaches are, thus, orthogonal:
our approach avoids the negative consequences of the scaling chain effect, but it just passively waits for the triggering event (significant increment in the inbound workload). The integration of machine learning techniques with our approach could further soften the impact of such an event leading to a better Quality of Service (e.g.\ latency and message loss).
%

Concerning future work, besides realizing the above described integration, we plan to
improve system simulation by accounting for failures (e.g., network partitioning, computing hardware failures) and their impact on the deployed system. 
To this aim, we could
evaluate the system following the practice of Chaos Engineering \cite{chaos-eng}, simulating the failures in ABS and making sure that the available resources are enough to guarantee a given level or robustness and resilience. 
Moreover, to improve the portability of our approach, we also plan to base our system modeling using a workflow language/notation that also includes data flow besides standard control flow, such as BPMN~\cite{bpmnSTD}. This will make it possible to automatically calculate microservice MCL and Multiplicative Factor according to formulae such as those used in our case study.

\bibliographystyle{abbrv}
\bibliography{biblio}

\appendix

\section{Appendix}

\subsection{SmartDeployCost Annotation Example}\label{SmartDeployCost}

Below we present the JSON description in the $SmartDeployCost$ annotation of a microservice class, taken from our case study.
\begin{lstlisting}
{ "class": "MessageReceiver_LoadBalancer",
  "scenarios": [{
    "name": "default",
    "provide": -1,
    "cost": {"Cores": 2,"Memory": 200},
    "sig":[{"kind":"require","type":"DBInterface"}],
    "methods": [{
         "add": {
              "name": "connectInstance", 
              "param_type": "MessageReceiverInterface"},
          "remove": {
              "name": "disconnectInstance", 
              "return_type": "MessageReceiverInterface"},
    }]
}]}
\end{lstlisting}
The keyword \textit{class} declares the name of the class which the annotation refers to and the keyword \textit{scenarios} contains a list of the possible deployment modalities (we just use the ``default'' one), each of them specifying a different set of requirements for the class. Such requirements are: in the \textit{provide} field, the number of objects that can use the ports (methods) provided by an object of the class, where $-1$ states that object ports can be used without restrictions; in the \textit{cost} field, the resources consumed by an object of the class;  in
the \textit{sig} field, the classes of the reference paramaters to be supplied to the class constructor (declaration that the class strongly requires ports of such classes); finally, in the \textit{methods} field, the class method names that can be used 
to add or remove additional references of a certain class (the class weakly requires ports of such class).


\subsection{Local Adaptation Algorithm}\label{localAdaptation}

In the local adaptation algorithm, each microservice (type) has a dedicated monitor and it is locally replicated by creating new instances every time its monitor detects that scaling is needed. The monitor code excerpt below, which is periodically executed (e.g.\ every 10 seconds in our case study), works as follows. First it applies the above described scale up/scale down conditions, with the constant $\mathsf{mcl}$ being the microservice MCL and the variable $\mathsf{deployedInstances}$ the number of deployed instances. Such a variable is assumed to be initially set to the value $\mathsf{baseInstanceN}$, i.e.\ the number of instances that the microservice has in the \textbf{B} configuration
(see Table \ref{table:tab2} in our case study), and is updated by the code in case of scaling.
Then it computes the minimum number of microservice instances needed to handle the incoming workload as $\mathsf{\lceil(inbound\_workload + K) / MCL\rceil}$. Finally it deploys/undeploys instances so to reach such a calculated optimal number. In particular, the method \textit{deploy($\dots$)}, besides incrementing the number of instances, it also dynamically modifies VMs speed according to the logic followed in Section \ref{timedSD}. If scale down occurs, the system keeps installed at least $\mathsf{baseInstanceN}$ instances.
\begin{lstlisting}
if((inbound_workload+kbig)-(mcl*deployedInstances)>k || %*\newline*) (mcl*deployedInstances)-(inbound_workload+kbig)>k) {
 Int configurationInstances = ceil(float((inbound_workload+kbig)/mcl));
 if(configurationInstances>deployedInstances) {
   s!deploy(configurationInstances-deployedInstances);}
 else if(configurationInstances<deployedInstances && deployedInstances>=baseInstanceN) {
   s!undeploy(deployedInstances-configurationInstances);}
 deployedInstances = configurationInstances;}
\end{lstlisting}

\end{document}